\documentclass[aps,pra,twocolumn,amsmath,showpacs,letterpaper,floatfix]{revtex4-1}

\usepackage{amssymb}
\usepackage{graphicx}
\usepackage{grffile}
\usepackage{color}

\begin{document}

\title{Classical and quantum-mechanical phase space distributions}
\author{Thomas Kiesel}
\affiliation{Arbeitsgruppe Quantenoptik, Institut f\"ur Physik, Universit\"at  Rostock, D-18051 Rostock,
Germany}
\begin{abstract}
	We examine the notion of nonclassicality in terms of quasiprobability distributions. In particular, we do not only ask if a specific quasiprobability can be interpreted as a classical probability density, but require that characteristic features of classical electrodynamics are resembled. We show that the only quasiprobabilities which correctly describe the superposition principle of classical electromagnetic fields are the $s$-parameterized quasiprobabilities. Furthermore, the Glauber-Sudarshan $P$ function is the only quantum-mechanical quasiprobability which is transformed at a classical attenuator in the same way as a classical probability distribution. This result strengthens the definition of nonclassicality in terms of the $P$ function, in contrast to possible definitions in terms of other quasiprobabilities.
\end{abstract}

\maketitle

\section{Introduction}

Since the early days of quantum optics, many effects have been observed which are incompatible with the classical electromagnetic theory of light. Already Einstein's explanation of the photoelectric effect~\cite{Einstein1905}, for which the photon was postulated, was in contradiction to the classical theory and gave rise to quantum mechanics. Later, Schr\"odinger found quantum mechanical states of the harmonic oscillator, which mostly resembled classical  behavior~\cite{Schrodinger}. Today, these states are known as coherent states of the electromagnetic field. Subsequently, much work has been done to understand the difference between the classical electromagnetic field and the quantum optical counterpart.

In the 1960s, physicists began to treat this problem in terms of phase space distributions. Classically, any state of a single mode of light can be described by a classical probability distribution over complex field amplitudes $\alpha$, which span the so-called  phase space. Quantum mechanically, phase-space distributions are not uniquely defined. Well-known examples are the Wigner distribution~\cite{Wignerfunction}, the Glauber-Sudarshan $P$ function~\cite{SudarshanP,GlauberP}, and the Husimi $Q$ function~\cite{HusimiQ}. They have been generalized as members of the class of $s$-parameterized quasiprobability distributions~\cite{CahillGlauber}. The most general form of quasiprobabilities has been given by Agarwal and Wolf~\cite{AgarwalWolf}. 

Nonclassicality can be defined by comparing such a quasiprobability with a classical probability distribution. However, this definition depends on the particular choice of quasiprobability. The generally accepted definition of nonclassicality has been formulated by Titulaer and Glauber: Whenever the  Glauber-Sudarshan $P$ function cannot be interpreted as a classical probability density, i.e.~has negativities in some well-defined sense, the state is referred to as nonclassical~\cite{TitulaerGlauber, MandelNonclassicality}. This definition is designed in such a way that any correlation function of a classical state can be modelled by classical electrodynamics. Furthermore, the coherent states, which we already mentioned above, are the only pure quantum mechanical states which are classical in this sense~\cite{Cahill,Hillery}. However, the $P$ function may be highly singular, such that it cannot be directly reconstructed from experimental data. This fact imposes severe difficulties to test nonclassicality of a state in general. Therefore, many nonclassicality criteria had to be developed, such as squeezing~\cite{Squeezing}, sub-Poissonian photon statistics~\cite{SubPoisson}, inequalities in terms of matrices of moments~\cite{AgarwalTara,Shchukin1,Shchukin2} or characteristic functions~\cite{Vogel2000,Vogel2002}, as well as bounds on probabilities~\cite{Luis}. The most general, but still simple criterion is based on so-called nonclassicality quasiprobabilities~\cite{NonclassicalityQuasiprobabilityTheory}, which has already been applied in practice~\cite{NonclassicalityQuasiprobabilitySPATS,NonclassicalityQuasiprobabilitySqueezed} and connected to nonclassicality witnesses~\cite{NonclassicalityWitnesses,NonclassicalityWitnessesMeasurement}.

A frequently used criterion is based on  the Wigner function. Its negativities are sufficient for nonclassicality in the sense of negative $P$ functions, but not necessary~\cite{WignerNegativities}. Moreover,  it can be obtained by standard means of quantum tomography, and therefore contains full information about the quantum state. Therefore, it is readily observed in many quantum optics laboratories and used for nonclassicality tests~\cite{Wineland,OpticalPhotonWigner,Silberhorn}.  Maybe it has been this  development which led to the proposal to define nonclassicality in terms of the Wigner function~\cite{Eisert}. Naturally the question arises whether this new definition is more suitable than the one based on the $P$ function.

In this article, we strengthen the role of the $P$ function as the key quantity for the definition of nonclassicality. The foundation of our considerations is the superposition principle of the classical electromagnetic field of light, which is employed at every beam splitter in an optical experiment. We demonstrate that not every general quasiprobability is transformed by a beam splitter in the same way as a classical phase-space distribution. More precisely, only the $s$-parameterized quasiprobabilities behave classically in this situation. Hence they are the only candidates for a reasonable definition of nonclassicality. Furthermore, we show that only the Glauber-Sudarshan $P$ function behaves under the action of an attenuator like a classical phase-space distribution. Hence, it is the only quasiprobability which completely resembles the classical behavior of linear operations on the classical field of light. This result supports the definition of  nonclassicality in terms of negativities of the $P$ function, notwithstanding its mathematical complications. Furthermore, it sheds a light on physical problems which will occur when nonclassicality shall be defined by negativities of the Wigner function. 

The paper is structured as follows: In Sec.~II, we compare the action of a beam splitter on classical phase space distributions and quantum-mechanical quasiprobabilities. In Sec.~III, the attenuator is analyzed for classical and quantum optics. In Sec.~IV, we give some explicit examples  to demonstrate that only the $P$ function can distinguish between classical and nonclassical behavior correctly, and show that a state with a nonnegative Wigner function may lead to measurement results which cannot be explained by classical electrodynamics. 

\section{Action of a beam splitter}
\subsection{The classical superposition principle}

Let us consider the electric field of a monochromatic light wave with a fixed frequency $\omega$. The spatial dependence of the field shall be neglected in this manuscript. Then, the field can be fully characterized by a complex amplitude $\alpha$ 
\begin{equation}
	E(\tau;\alpha) = \alpha e^{i\omega \tau} + \alpha^* e^{-i\omega \tau}.
\end{equation}
The maximum amplitude is given by $E_{\rm max}(\alpha) = 2 {\rm Re}(\alpha)$, and the phase can be specified by $\arg(\alpha)$. The set of all complex $\alpha$ is referred to as phase space.

\begin{figure}
	\includegraphics[width=0.6\columnwidth]{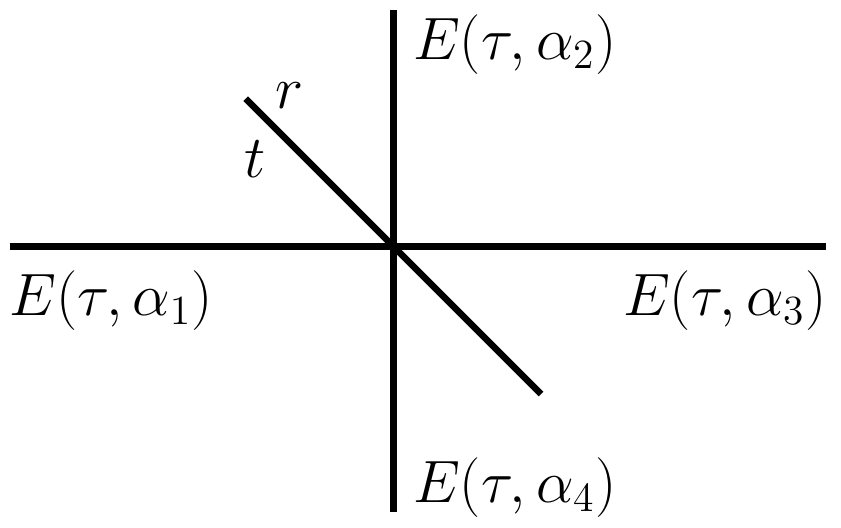}
	\caption{Scheme of a beam splitter.}
	\label{fig:beamsplitter}
\end{figure}

A characteristic feature of light is the possibility of linear superposition of electromagnetic fields, which can be achieved by a beam splitter, see Fig.~\ref{fig:beamsplitter}. The relation of input and output fields $E(\tau,\alpha_i)$ are given by
\begin{equation}
	\left(\begin{array}{c} E(\tau,\alpha_3)\\E(\tau,\alpha_4)\end{array}\right)
		=   U \left(\begin{array}{c} E(\tau,\alpha_1)\\E(\tau,\alpha_2)\end{array}\right),\label{eq:beamsplitter:field}
\end{equation}
where the matrix 
\begin{equation}
	U = \left(\begin{array}{c c} t & r\\ -r^* & t^*\end{array}\right) e^{i\varphi_U}\label{eq:beamsplitter:matrix}
\end{equation}
has to be a unitary one. Throughout the manuscript, we set the global phase $\varphi_U = 0$. Instead of considering the fields, this map can also be considered in terms of the complex amplitudes $\alpha_i$, 
\begin{equation}
	\left(\begin{array}{c}\alpha_3\\\alpha_4\end{array}\right)
		=   U	\left(\begin{array}{c} \alpha_1\\\alpha_2\end{array}\right).\label{eq:classical:beamsplitter}
\end{equation}
If all parameters are non-zero, the knowledge of two arbitrary amplitudes $\alpha_i,\alpha_j$ complete determines the missing two amplitudes. This is trivial if the amplitude $\alpha_1,\alpha_2$ or $\alpha_3,\alpha_4$ are given.  If $\alpha_1, \alpha_3$ are known, the missing two fields are determined by. 
$$
	\left(\begin{array}{c} \alpha_2\\\alpha_4\end{array}\right)
		=   \frac{1}{r}\left(\begin{array}{c c} -t & e^{-i\varphi_U}\\ -e^{i\varphi_U} & t^*\end{array}\right)
			\left(\begin{array}{c} \alpha_1\\\alpha_3\end{array}\right).$$
All other cases can be treated analogously. Therefore, whenever two classical fields $E(t,\alpha_i)$, $E(t,\alpha_j)$ are given, there exist two additional classical fields such that Eq.~\eqref{eq:beamsplitter:field} is satisfied.

Moreover, we may consider a statistical ensemble of fields, modelled by a probability distribution $P_{\rm cl.}(\alpha)$. In this situation, every physical quantity $F(\alpha)$ has to be averaged over this probability distribution,
\begin{equation}
	\langle\hat F\rangle  = \int d^2\alpha\, P_{\rm cl.}(\alpha) F(\alpha).
\end{equation}
The action of a beam splitter on a two-mode input field characterized by its joint probability density $P_{12}(\alpha_1,\alpha_2)$ is given by
\begin{equation}
	P_{34}(\alpha_3,\alpha_4) = P_{12}(t^*\alpha_3-r\alpha_4, r^*\alpha_3 + t\alpha_4)\label{eq:beamsplitter:on:Pcl}.
\end{equation}
Therefore, for each classical input field ensemble there exists a classical output field, defined by the probability distribution $P_{34}(\alpha_3,\alpha_4)$. Due to the linearity of the relations, this holds for an arbitrary choice of two given modes $i,j$.

\subsection{The beam splitter and quantum phase-space functions}

Quantum mechanically, the expectation value of an arbitrary operator $\hat F$ can be calculated in the form \cite{AgarwalWolfII}
\begin{equation}
	\langle\hat F \rangle = \int d^2\alpha\, P_\Omega(\alpha) F_{\tilde \Omega}(\alpha).
\end{equation}
Here, the so-called quasiprobability $P_\Omega(\alpha)$ is given by the Fourier transform of its characteristic function $\Phi_\Omega(\beta)$~(cf.~\cite{AgarwalWolf}),
\begin{equation}
	P_\Omega(\alpha) = \frac{1}{\pi^2}\int d^2{\beta}\,\Phi_\Omega(\beta) e^{\beta^*\alpha-\beta\alpha^*},
\end{equation}
which itself is directly connected to the quantum state of the oscillator, described by its density operator $\hat \rho$:
\begin{equation}
	\Phi_\Omega(\beta) = {\rm Tr}(\hat \rho  e^{\beta\hat a^\dagger-\beta^*\hat a}) \Omega(\beta,\beta^*).
\end{equation}
The filter function $\Omega(\beta)$ characterizes the special kind of the quasi-probability, which we examine more in detail below. The function $F_{\tilde \Omega}(\alpha)$ is connected to the operator $\hat F$ in a similar way, with $\tilde \Omega(\alpha)$ being a suitable inverse filter. 

The different known types of quasiprobabilities are determined by a specific filter $\Omega(\beta)$. For instance, the Wigner function can be obtained by $\Omega(\beta) = 1$, the $P$ function by $\Omega(\beta) = e^{|\beta|^2/2}$ and the Husimi $Q$ function by $\Omega(\beta) = e^{-|\beta|^2/2}$. The latter quasiprobabilities are generalized as $s$-parameterized quasiprobabilities in \cite{CahillGlauber}, where the filter has the form $\Omega(\beta)= e^{s|\beta|^2/2}$. In general, we require that $\Omega(\beta)$ is an analytic function in $\beta$ and $\beta^*$, has no zeros, and satisfies $\Omega(0) = 1$. Under this conditions, we derive our first important result:

{\bf Theorem: } 
	Only the s-parameterized quasiprobabilities resemble the classical action of a beam 	splitter.

\emph{Proof:} It appears that the action of the beamsplitter on ensembles of electromagnetic fields  can be examined in a convenient way in terms of the characteristic function $\Phi_{34}(\beta_3,\beta_4)$ of the probability distribution~\eqref{eq:beamsplitter:on:Pcl}, which is defined as
\begin{align}
	\Phi_{34}(\beta_3,\beta_4) =& \int d^2\alpha_3\int d^2\alpha_4\,P_{34}(\alpha_3,\alpha_4) \nonumber\\&\qquad\times e^{\beta_3\alpha_3^* + \beta_4\alpha_4^* - \beta_3^*\alpha_3 - \beta_4^*\alpha^4} .
\end{align}
Then Eq.~\eqref{eq:beamsplitter:on:Pcl} can be rewritten in terms of characteristic functions as
\begin{equation}
	\Phi_{34}(\beta_3,\beta_4) = \Phi_{12}(t^*\beta_3 - r\beta_4, r^*\beta_3 + t\beta_4).
	\label{eq:beamsplitter:on:charfunc}
\end{equation}
In the following, we seek for the characteristic functions $\Phi_\Omega(\beta_3,\beta_4)$ of the quasiprobabilities which also satisfy this equation. The theorem states that only quasiprobabilities with $\Omega(\beta) = e^{s|\beta|^2/2}$ fulfill this requirement.

To show this, we first note that in the case of two-mode fields, the filter function of the two-mode characteristic function  $\Phi_\Omega(\beta_3,\beta_4)$ has to appear as a product of single mode filter functions,
\begin{equation}
\Phi_\Omega(\beta_3,\beta_4) = {\rm Tr}(\hat \rho  e^{\beta_3\hat a_3^\dagger + \beta_4\hat a_4^\dagger-\beta_3^*\hat a_3-\beta_4^*\hat a_4})\Omega(\beta_3)\Omega(\beta_4).
\end{equation}
Then, the action of the beamsplitter on the creation and annihilation operators can be written analogously to Eq.~\eqref{eq:classical:beamsplitter}, cf.~\cite{KlauderBeamsplitter}:
\begin{equation}
	\left(\begin{array}{c}\hat a_3\\\hat a_4\end{array}\right)
		=   U	\left(\begin{array}{c} \hat a_1\\\hat a_2\end{array}\right).\label{eq:beamsplitter}
\end{equation}

Let us denote the characteristic function of the Wigner function, defined by $\Omega(\beta) \equiv 1$, as $\Phi_W(\beta_3,\beta_4)$. We can easily show that  Eq.~\eqref{eq:beamsplitter:on:charfunc} holds for $\Phi_W(\beta_3,\beta_4)$. Therefore, it is sufficient to examine the relation of the beamsplitter on the filter function $\Omega(\beta)$. More precisely, any characteristic function $\Phi_\Omega(\beta_3,\beta_4)$ satisfies the classical relation Eq.~\eqref{eq:beamsplitter:on:charfunc} if and only if the filter function $\Omega(\beta)$ satisfies
\begin{equation}
	\Omega(\beta_3)\Omega(\beta_4) = \Omega(t^*\beta_3 - r\beta_4)\Omega( r^*\beta_3 + t\beta_4). \label{eq:beamsplitter:condition:Omega}
\end{equation}
The only conditions which are imposed on the beamsplitter coefficients $t, r$ arise from the unitarity of the matrix $U$ in Eq.~\eqref{eq:beamsplitter:matrix}:
\begin{equation}
	\quad |t|^2 + |r|^2 = 1.\label{eq:beamsplitter:relations}
\end{equation}
All filter functions $\Omega(\alpha)$ which are entire in $\alpha,\alpha^*$ and  have no zeros, can be written in the form (cf.~\cite{AgarwalWolf})
\begin{equation}
 \Omega(\beta) = \exp\left\{\sum_{k,l=0}^\infty c_{kl} \beta^k(\beta^*)^l\right\}.
\end{equation}
We may insert this expression into Eq.~\eqref{eq:beamsplitter:condition:Omega} and compare the coefficients of the series in the exponential functions. 

Since we require $\Omega(0) = 1$, we need $c_{00} = 0$. Let us now examine all powers of $\beta_3^k(\beta_3^*)^l$ with $k + l > 0$. Comparing both sides of the equation, we find the condition 
\begin{equation}
	c_{kl} = c_{kl} [(t^*)^k t^l + (r^*)^k r^l].
\end{equation}
The equation is fulfilled for a fixed pair $k,l$ if the expression in brackets equals to $1$ for all allowed parameters $r, t$, or if $c_{kl} = 0$. It is easy to show that two special cases already lead to a large number of vanishing coefficients:
\begin{itemize}
	\item If we insert the valid parameters $t = r=1/\sqrt{2}$, the term in brackets is given by
		\begin{equation}
			[(t^*)^k t^l + (r^*)^k r^l] = 2/\sqrt{2^{k+l}}
		\end{equation}
		This expression can only be equal to one if $k + l = 2$. Therefore, it is necessary that $c_{kl} = 0$ for $k+l > 2$ as well as $c_{10} = c_{01} = 0$.
	\item If we insert the valid parameters $t = 1/\sqrt{2}$ and $r = i/\sqrt{2}$, and consider the case $k + l = 2$ only, the term in brackets is given by
		\begin{equation}
			[(t^*)^k t^l + (r^*)^k r^l] = (1 - (-1)^k)/2,
		\end{equation}
		which equals to $1$ if and only if $k = l = 1$. Therefore, we find $c_{20} = c_{02} = 0$.
\end{itemize}
Therefore, after setting $c_{11} = s/2$ we find that only filters of the form 
\begin{equation}
	\Omega(\beta) = \exp\{s |\beta|^2/2\} \label{eq:Omega:s}
\end{equation}	
are candidates to satisfy Eq.~\eqref{eq:beamsplitter:condition:Omega}. 

As a last step, we still have to show that the filter function~\eqref{eq:Omega:s} fulfills Eq.~\eqref{eq:beamsplitter:condition:Omega}. To this end, we start with the right hand side of the latter equation:
\begin{align}
	&\Omega(t^*\beta_3 - r\beta_4)\Omega( r^*\beta_3 + t\beta_4)\nonumber\\
		&= \exp\left\{s[(|t|^2+|r|^2)|\beta_3|^2 + (|r|^2+|t|^2)|\beta_4|^2\right.\nonumber\\
		&\qquad \left.+ (r t-t r)^*\beta_3\beta_4^* + (r t-t r)\beta_3^*\beta_4]/2\right\}.
\end{align}
Now, we can insert the relations~\eqref{eq:beamsplitter:relations} and easily verify that
\begin{align}
	&\Omega(t^*\beta_3 - r\beta_4)\Omega( r^*\beta_3 + t\beta_4)\nonumber\\
		&\qquad\qquad= \exp\left\{s[|\beta_3|^2 +|\beta_4|^2]/2\right\}.
\end{align}
The last term exactly equals to the left hand side of the desired equality, 
cf.~Eq.~\eqref{eq:beamsplitter:condition:Omega}. This concludes the proof.\hfill$\Box$

\section{Attenuation of the electromagnetic field}

So far, we have shown that the action of a beam splitter on quantum mechanical quasiprobability distributions can only resemble the classical behavior if the quasiprobabilities belong to a special class, namely the s-parameterized ones. Hence, as the $Q$-function, Wigner function and $P$ function all belong to this class, they are suitable for a 
classical description of the beam splitter action. However, there are $s$-parameterized quasiprobability distributions which are still not suitable for the definition of nonclassicality. For instance, the Husimi $Q$ function always satisfies the requirements for a classical probability density.

In this section, we will impose an additional requirement. Classically, damping of the electromagnetic field can be considered as the action on the beam splitter, where a signal field $\alpha_1$ is overlapped with the vacuum field $\alpha_2 = 0$, leading to a resulting attenuated field $\alpha_3 = t \alpha_1$. The parameter $\eta = |t|^2$ is commonly referred to as (quantum) efficiency, which diminishes the amplitude of the signal field. The joint probability density of the input field is given by the product of the probability densities of the state of interest and the vacuum field,
\begin{equation}
	P_{12}(\alpha_1,\alpha_2) = P_{1}(\alpha_1) \delta(\alpha_2).
\end{equation}
The probability density of the attenuated output field $\alpha_3$ is then obtained as the marginal of $P_{34}(\alpha_3,\alpha_4)$, given by Eq.~\eqref{eq:beamsplitter:on:Pcl}:
\begin{eqnarray}
	P_{3}(\alpha_3) &=& \int d^2\alpha_4 P_{1}(t^*\alpha_3-r\alpha_4)\delta(r^*\alpha_3+t\alpha_4) \nonumber \\
 &=& \frac{1}{|t|^2}P_{1}(\alpha_3/t)\label{eq:attenuator:classical:P}
\end{eqnarray}
Vice versa, given an classical output field distribution $P_3(\alpha_3)$ and the quantum efficiency $\eta = |t|^2 >0$, there always exists a classical distribution $P_1(\alpha_1) = |t|^2 P_3(\alpha_1 t)$  for a classical input field of the attenuation process. 

Quantum-mechanically, attenuation is also considered as overlapping the signal field with vacuum at the beam splitter, described by quasiprobabilities $P_{\Omega,1}(\alpha_1)$ and  $P_{\Omega, \rm vac} (\alpha_2)$. Therefore, the input state is described by the quasiprobability
\begin{equation}
	P_{\Omega,12}(\alpha_1,\alpha_2) = P_{\Omega,1}(\alpha_1) P_{\Omega, \rm vac} (\alpha_2).\label{eq:attenuator:P:Omega}
\end{equation} 
This leads us to the following result:

{\bf Theorem:}
	The Glauber-Sudarshan $P$ quasiprobability is the only quasiprobability which allows to describe the action of an attenuator in the same way as the classical electromagnetic theory, see Eq.~\eqref{eq:attenuator:classical:P}.

\emph{Proof:} Again, we examine the quasiprobabilities in terms of their characteristic functions. The classical relation~\eqref{eq:attenuator:classical:P} can be rewritten as
\begin{equation}
\Phi_{3}(\beta_3) = \Phi_{1}(t^*\beta_3)\label{eq:attenuator:CF:classical}.
\end{equation}
On the other side, we have to insert the Fourier transform of Eq.~\eqref{eq:attenuator:P:Omega} into the quantum-mechanical version of Eq.~\eqref{eq:beamsplitter:on:charfunc}:
\begin{equation}
\Phi_{\Omega,34}(\beta_3,\beta_4) = \Phi_{\Omega,1}(t^*\beta_3 - r\beta_4) \Phi_{\Omega,\rm vac}( r^*\beta_3 + t\beta_4).
\end{equation}
Since we consider only the output mode $3$, we have to take the marginal with respect to the mode $4$. In Fourier space, this is simply achieved by setting $\beta_4 = 0$. Therefore, the output mode $3$ is described by the characteristic function
\begin{equation}
\Phi_{\Omega,3}(\beta_3) = \Phi_{\Omega,1}(t^*\beta_3) \Phi_{\Omega,\rm vac}( r^*\beta_3)\label{eq:attenuator:CF:Omega}.
\end{equation}
Comparing this result with the classical expectation in Eq.~\eqref{eq:attenuator:CF:classical}, we find 
\begin{equation}
	\Phi_{\Omega,\rm vac}( \beta) \equiv 1
\end{equation}
as the necessary and sufficient condition for the equality of the quantum mechanical and classical quasiprobabilities. This is only satisfied for the characteristic function of the $P$ function, which is obtained by $\Omega(\beta) = e^{|\beta|^2/2}$. \hfill$\Box$

We emphasize that the key argument of the proof is that the quasiprobability for the vacuum state exactly equals to the classical probability distribution for a field with zero amplitude, namely $P_{\rm vac}(\alpha) = \delta(\alpha)$.

\section{Discussion}

\subsection{Second-order correlation measurement}
\begin{figure}
	\includegraphics[width=0.6\columnwidth]{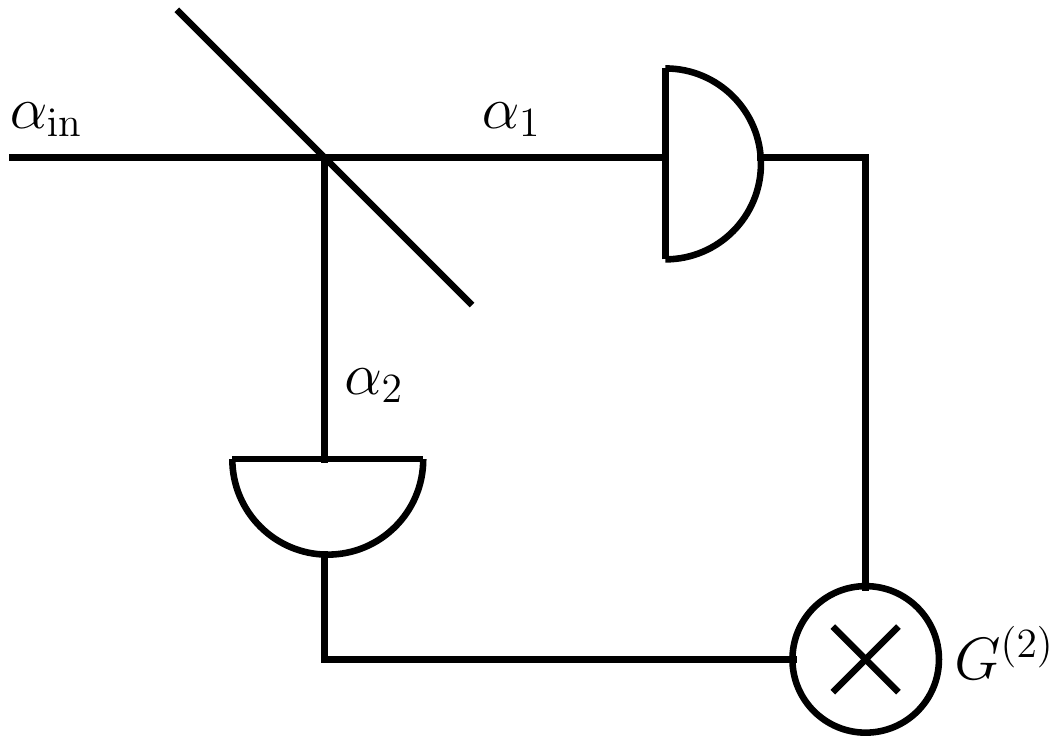}
	\caption{Experimental setup of second-order correlation measurements. The input field is split by a symmetric beamsplitter. The intensities of the output field are correlated.}
	\label{HBT-Setup}
\end{figure}
First, let us illustrate our conclusions with a simple example. For this purpose, we consider the second-order correlation measurement (Fig.~\ref{HBT-Setup}), as it is used for photon-antibunching experiments~\cite{PhotonAntibunching}. In classical terms, the input field $\alpha_{\rm in}$ is split by the symmetric beam splitter into two output fields $\alpha_{1,2} \propto \frac{\alpha_{\rm in}}{\sqrt{2}}$, whose intensity correlation is measured,
\begin{equation}
	G^{(2)}_{\rm cl.}(\alpha_{\rm in})  = |\alpha_{1}|^2 |\alpha_{2}|^2 \propto  |\alpha_{\rm in}|^4.
\end{equation} 
Consequently, for any statistical superposition of input fields $\alpha_{\rm in}$, the correlation function reads as
\begin{equation}
	G^{(2)}_{\rm cl.} = \int P_{\rm cl.}(\alpha) G^{(2)}_{\rm cl.}(\alpha)  d^2\alpha
\end{equation}
Moreover, one may show that any classical field satisfies the inequality\cite{TitulaerGlauber, VogelPRL100}
\begin{equation}
	G^{(2)}_{\rm cl.}  \geq [G^{(1)}]^2,\label{eq:classical:inequality}
\end{equation}
with the first order correlation function 
\begin{equation}
	G_{\rm cl.}^{(1)} = 	\int P_{\rm cl.}(\alpha) |\alpha|^2d^2\alpha,
\end{equation}
describing the intensity of the input field.

As a first quantum mechanical system, let us consider a single photon, which has a Wigner function with negativities as well as a highly singular $P$ function. Hence, in terms of both quasiprobabilities, the photon can be referred to as nonclassical. We obtain the  correlation functions as
\begin{equation}
	G^{(1)} \propto \langle\hat a^\dagger\hat a\rangle = 1
\end{equation}
and
\begin{equation}
	G^{(2)} \propto \langle\hat (a^\dagger)^2\hat a^2 \rangle = 0.
\end{equation}
Obviously, the photon does not satisfy inequality~\eqref{eq:classical:inequality}. Therefore, the outcomes of such an experiment cannot be explained by means of classical electrodynamics, and the term "nonclassicality" is justified.

\begin{figure}
	\includegraphics[width=0.8\columnwidth]{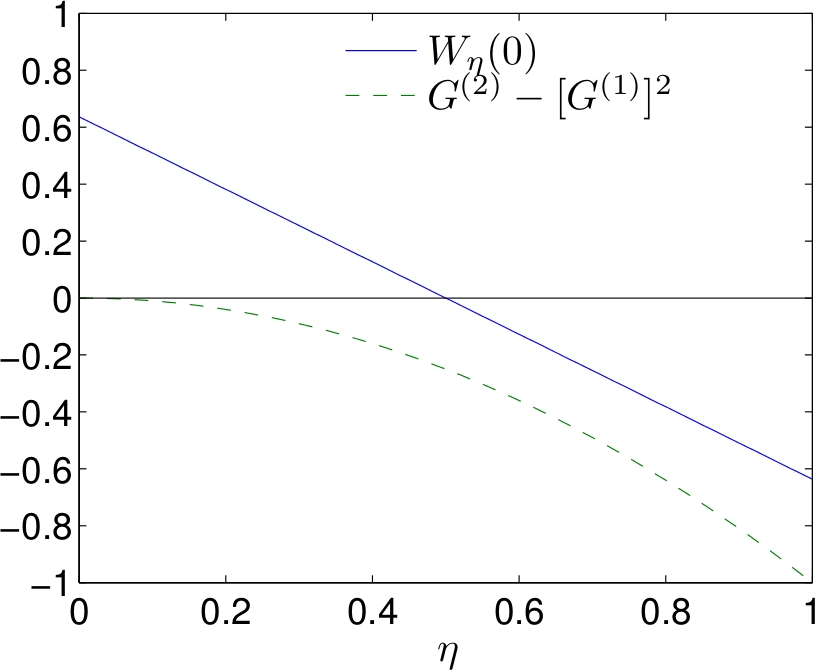}
	\caption{(Color online) Wigner function at the origin of phase space (solid blue line) and correlation functions (green dashed line) of a single photon, undergoing an attenuation of $\eta$. Negativity of each quantity indicates nonclassicality. Clearly, the Wigner function does not indicate nonclassicality for $\eta \leq 1/2$, although the correlation functions violate the classical boundary.}
	\label{fig:Wigner:photon}
\end{figure}

However, the situation changes when we examine a slightly different quantum state, namely a single photon, which is attenuated by a value of $\eta = 1/2$. Classically, the intensity is simply divided by two. Quantum-mechanically, the field has to be described by effective  annihilation operators 
\begin{equation}
	\hat b = (\hat a + \hat a_{\rm vac})/\sqrt{2},
\end{equation}
where $\hat a$ is the annihilation operator of the signal field and $\hat a_{\rm vac}$ is the annihilation operator of an auxiliary vacuum field. In this case, the correlation functions are given by
\begin{equation}
	G^{(1)} \propto \langle\hat b^\dagger \hat b\rangle = \frac{1}{2}
\end{equation}
and 
\begin{equation}
	G^{(2)} \propto \langle\hat (b^\dagger)^2\hat b^2 \rangle = 0.
\end{equation}
Therefore, the classical inequality \eqref{eq:classical:inequality} is still violated. However, the Wigner function of the input state is given by~\cite{NonclassicalitySPATS}
\begin{equation}
	W_\eta(\alpha) = \frac{2}{\pi} [1 -2\eta + 4\eta |\alpha|^2]e^{-2|\alpha|^2},
\end{equation}
being nonnegative for $\eta \leq 1/2$. Hence, in terms of the Wigner function, one might state that the quantum state is classical, although the second-order correlation function violates basic laws of classical electrodynamics. This is also illustrated in Fig.~\ref{fig:Wigner:photon}: For any positive $\eta$, the classical bound on the correlation functions is violated, while the Wigner function only has negativities for $\eta > 1/2$.  As we have shown in the previous section, the $P$ function is the only quasiprobability which resembles the classical behavior of the attenuator, and is free of such a contradiction.

\subsection{General correlation functions}

In a more general setup, complex moments of the electromagnetic field can be analyzed, 
\begin{equation}
	G_{\rm cl.}^{(m,n)} = 	\int P_{\rm cl.}(\alpha) (\alpha^*)^{m} \alpha^n d^2\alpha.
\end{equation}
Since the field amplitude $\alpha$ scales with the square root of the attenuation $\eta$, the moments of the attenuated field are given by
\begin{equation}
	G_{\rm cl.,\eta}^{(m,n)} = \eta^{\frac{m+n}{2}} G_{\rm cl.}^{(m,n)}.\label{eq:G:mn:eta}
\end{equation}
Due to this fact, inequalities for the correlation functions which set bounds on classical states appear to be independent of the efficiency. For instance, Eq.~(5.6) in~\cite{TitulaerGlauber} states that for classical fields the condition 
\begin{equation}
	G_{\rm cl.,\eta}^{(n,n)} \geq G_{\rm cl.,\eta}^{(m,m)}G_{\rm cl.,\eta}^{(n-m,n-m)}
\end{equation}
holds for all $n\geq m \geq 0$.  Inserting Eq.~\eqref{eq:G:mn:eta} directly shows that the latter inequality is satisfied if and only if it is also satisfied for the unattenuated state with $\eta = 1$. 

Even more generally, this also holds for field correlation functions measured at different points in space. Most generally, the complete hierarchy of nonclassicality criteria given in~\cite{VogelPRL100}, based on determinants of moments, is independent of the efficiency, since $\eta$ appears with the same power in each term of any determinant and can be factorized out. Although the latter result has been derived for states with nonnegative $P$ function, it also holds for classical mixtures of classical electromagnetic fields.

In conclusion, classicality of a state does not change under the influence of positive efficiency $\eta$. Conversely, a nonclassical quantum state shall not be changed into a classical one by attenuation. This key feature is only present in the Glauber-Sudarshan $P$ representation of the state, since it is the only quasiprobability which resembles the same scaling as the classical phase space probability, see Eq.~\eqref{eq:attenuator:classical:P}. As a consequence, its negativities (i.e.~nonclassicality) are preserved under the action of beamsplitter and attenuator. This gives the $P$ function  its exceptional position in the discussion of nonclassicality, and demonstrates that it is reasonable to consider nonclassicality in the sense of Titulaer and Glauber in spite of all mathematical challenges.

\subsection{Comment on balanced homodyne detection}

Sometimes it is argued that negativities of the Wigner function can be used for the definition of nonclassicality, since it can be seen as the joint probability density for all quadrature measurements~\cite{VogelRisken}. The latter are implemented by  balanced homodyne measurements, which are commonly used in most laboratories  for quantum state reconstruction. As shown above, a photon undergoing an attenuation of $\eta=1/2$ would be argued to be classical, since the joint probability distribution of the quadrature distributions, i.e.~the Wigner function, is nonnegative. Indeed, the balanced homodyne measurement result can be explained also by the state of a classical electrodynamic field. However, the notion of classicality is restricted to this particular type of measurement, and hence does not only depend on the state itself. If the same state is subject to a  different measurement, such as the second order correlation measurement discussed above, discrepancies to the classical behavior appear which have not been immediately expected from inspection of the Wigner function. This restricts the usability of the Wigner function for defining nonclassicality to a subclass of special measurement setups. 

\section{Conclusions}

We have examined the action of linear devices, such as the beamsplitter and the attenuator, on quantum mechanical phase-space distributions and compared it with the situation in classical electrodynamics. Our goal was to find out which quasiprobabilities are able to resemble the classical behavior, in order to be useful for the definition of nonclassicality. We found that for the beamsplitter all $s$-parameterized quasiprobabilities, such as the $Q$ function, the Wigner function and the $P$ function reflect the classical action. However, the attenuator can only be described analogously to the classical picture in terms of the $P$ function. Therefore, the latter is the only quasiprobability which maps the state of light in the same way as classical linear devices do, preserving the classical properties of a field. Hence, the $P$ function possesses properties which are superior for the definition of nonclassicality.

\end{document}